\documentclass{PoS}
\usepackage{wrapfig}

\title{Conserved charge fluctuations with smaller-than-physical quark masses}

\ShortTitle{Conserved charge fluctuations}

\author{\speaker{Mugdha Sarkar}\thanks{This work was supported by the Deutsche Forschungsgemeinschaft (DFG, German Research Foundation) - project number 315477589 - TRR 211, the German Bundesministerium f{\"u}r Bildung und Forschung through Grant No. 05P2018 (ErUM-FSP T01) and the EU H2020-MSCA-ITN-2018-813942 (EuroPLEx).}\\
        Fakult\"at f\"ur Physik, Universit\"at Bielefeld, D-33615 Bielefeld, Germany.\\
        E-mail: \email{mugdha@physik.uni-bielefeld.de}}

\author{Olaf Kaczmarek\\
        Fakult\"at f\"ur Physik, Universit\"at Bielefeld, D-33615 Bielefeld, Germany.\\
        Key Laboratory of Quark \& Lepton Physics (MOE) and Institute of Particle Physics,
        Central China Normal University, Wuhan 430079, China.\\
        E-mail: \email{okacz@physik.uni-bielefeld.de}}

\author{Frithjof Karsch\\
        Fakult\"at f\"ur Physik, Universit\"at Bielefeld, D-33615 Bielefeld, Germany.\\
        E-mail: \email{karsch@physik.uni-bielefeld.de}}

\author{Anirban Lahiri\\
        Fakult\"at f\"ur Physik, Universit\"at Bielefeld, D-33615 Bielefeld, Germany.\\
        E-mail: \email{alahiri@physik.uni-bielefeld.de}}
        
\author{Christian Schmidt\\
        Fakult\"at f\"ur Physik, Universit\"at Bielefeld, D-33615 Bielefeld, Germany.\\
        E-mail: \email{schmidt@physik.uni-bielefeld.de}}

\abstract{We present results from calculations of conserved charge fluctuations in $(2+1)$-flavor QCD
using light quark masses in the range $m_s/80 \leq m_l \leq m_s/27$, with the strange quark mass ($m_s$) kept
fixed at its physical value. This corresponds to a Goldstone pion mass in the range $80$ MeV $\leq m_\pi \leq 140$ MeV.
The measurements have been done using HISQ fermion discretization and Symanzik improved gauge action.
We discuss the quark mass dependence of up to 6th order cumulants and present first results on the separation
of singular and regular contributions to these cumulants. From these results, we examine the nature of the chiral
phase transition and the variation of the curvature of the crossover line as we approach the chiral limit.}

\FullConference{37th International Symposium on Lattice Field Theory - Lattice2019\\
		16-22 June 2019\\
		Wuhan, China}

\begin{document}

\section{Introduction}
\begin{wrapfigure}{r}{0.4\textwidth}
  \begin{center}
    \includegraphics[width=.35\textwidth]{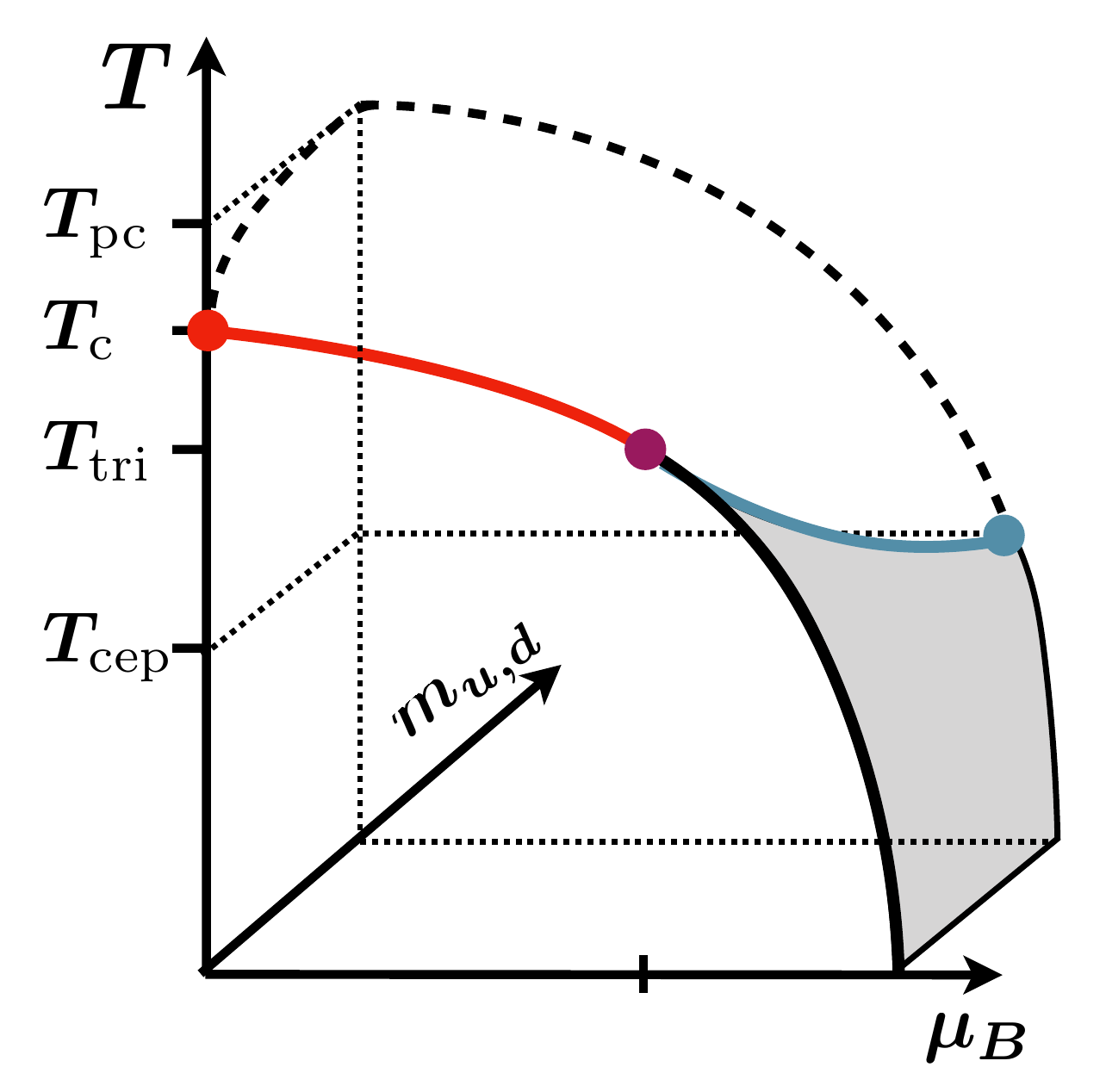}
  \end{center}
  \caption{Schematic QCD phase diagram \cite{Karsch:2019mbv} with temperature ($T$), baryon chemical potential ($\mu_B$) and light quark mass ($m_l$) axes.}
  \label{3dphasediag}
\end{wrapfigure}

The $SU(2)_L\times SU(2)_R$ chiral symmetry in QCD is explicitly broken due to non-zero up and down quark masses. At zero chemical potential with physical quark masses, it has been well established from lattice calculations that the chiral symmetry gets restored at high temperature through a crossover and not a phase transition. In Fig.~\ref{3dphasediag} \cite{Karsch:2019mbv}, we show a schematic phase diagram, inspired by various model calculations \cite{Halasz:1998qr,Stephanov:2006dn,Buballa:2018hux}. 
As seen in the figure, the crossover line, which continues for non-zero baryon chemical potential, is conjectured to end at a yet elusive critical point at temperature $T=T_\mathrm{cep}$. The nature of the different phase transitions in the QCD parameter space is crucial to our understanding of the phase diagram.

At zero chemical potential in the limit of two degenerate quark flavors becoming massless, the chiral transition is expected to become either a continuous phase transition belonging to 3d $O(4)$ universality class or a first order transition \cite{Pisarski:1983ms,Philipsen:2016hkv}. The possibility of a first order transition in the chiral limit would be associated with a $Z(2)$ end point at a finite quark mass. Recently in Ref. \cite{Ding:2019prx}, the chiral phase transition temperature $T_c$ (see Fig.~\ref{3dphasediag}) was determined to be $132^{+3}_{-6}$ MeV. It was shown that the chiral observables follow the volume and mass scaling consistent with $O(4)$ universality class. 

In this work, we study the fluctuations of conserved charges with lattice QCD at smaller-than-physical pion masses and zero chemical potential to understand the imprint of the critical phenomena at the chiral limit on these observables. Our preliminary results suggest that the fluctuations are consistent with $O(4)$ scaling expectations.  

\section{Observables and definitions}
According to renormalization group theory, near a phase transition, thermodynamic quantities have a universal contribution along with the theory-dependent regular terms. Depending upon the universality class, the universal part shows singular or non-analytic behavior very close to a phase transition, which gives rise to critical phenomena. It is useful to define generalised scaling variables corresponding to temperature and the symmetry-breaking mass variables, as
\begin{equation}
  t=\frac{1}{t_0}\left(\frac{T-T_c}{T_c} + \kappa_2^X\left(\frac{\mu_X}{T}\right)^2\right), \quad h=\frac{1}{h_0}\frac{m_l}{m_s}, \label{thdef}
 \end{equation}
 where $X$ stands for baryon number (B), electric charge (Q) or strangeness (S) conserved charges, with $\mu_X$ denoting the corresponding chemical potential and $\kappa^X_2$, the leading curvature coefficient of the transition line in the chiral limit. Here, $t_0$ and $h_0$ are dimensionless non-universal parameters. The chiral phase transition temperature at zero quark masses is denoted as $T_c$. Although the symmetry breaking variable is $m_l$, the bare mass of the two degenerate light fermions, it is convenient to define $h$ in terms of the ratio $m_l/m_s$, where $m_s$ is the bare mass of the strange quark.
 
Near the chiral phase transition ($m_l\to 0$), the free energy density of QCD at finite temperature and density, can thus be expressed as the sum of non-analytic and regular terms as
\begin{equation}
  f(T,\vec{\mu},m_l)/T^4 = h^{(2-\alpha)/\beta\delta}f_f(z) + f_r(T,\vec{\mu},m_l),
 \end{equation}
where $f_f(z)$ denotes a universal scaling function as a function of scaling variable $z\equiv t/h^{1/\beta\delta}$ with $\alpha, \beta$ and $\delta$ being critical exponents of the applicable universality class. The non-universal regular terms are denoted altogether as $f_r$. 

\begin{figure}
\centering
\includegraphics[width=.35\textwidth]{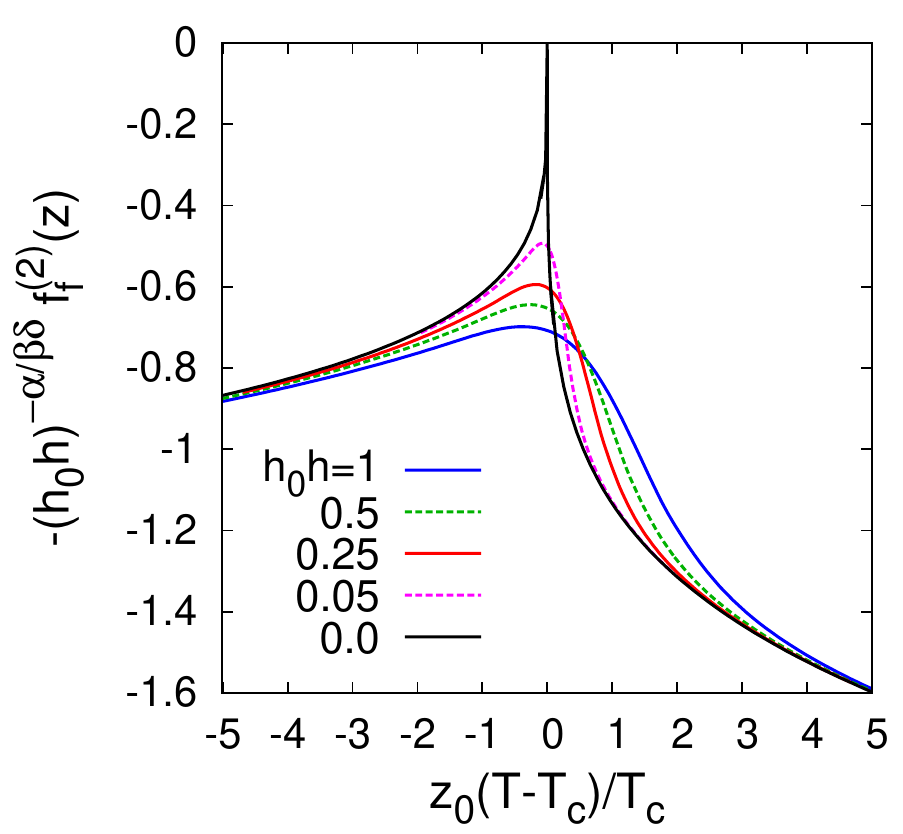}~~~~\quad
\includegraphics[width=.35\textwidth]{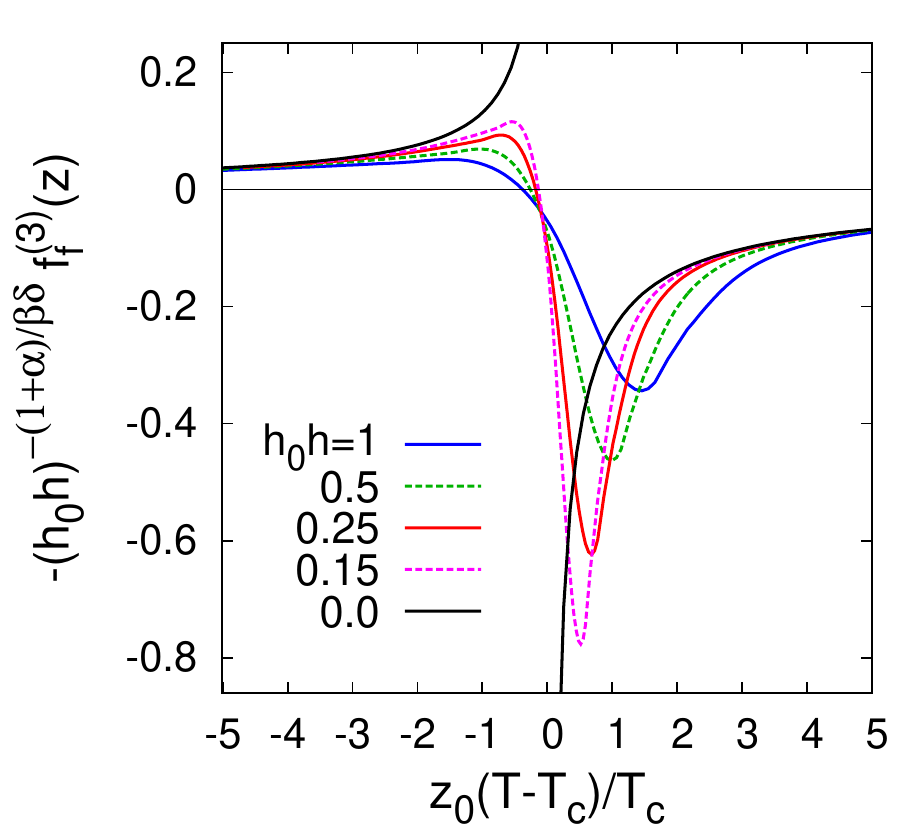}
\caption{Figures depicting the scaling behavior of the universal non-analytic contributions to $\chi_4$ (left) and $\chi_6$ (right), calculated using 3d $O(4)$ universality class scaling functions \cite{Friman:2011pf}. In both plots, $z_0=h_0^{1/\beta\delta}/t_0$ is a non-universal scale parameter.}
\label{o4plots}
\end{figure}

The fluctuations of the conserved charges with respect to chemical potential $\mu$ reveal important information about the critical behavior. At $\mu=0$, the singular part of such fluctuations can be obtained as \cite{Friman:2011pf}
\begin{equation}
 \chi_{2n}^X = -\frac{\partial^{2n}f/T^4}{\partial(\mu_X/T)^{2n}}\Bigg|_{\mu_X=0}\sim \quad -~(2\kappa^X_2)^n ~h^{(2-\alpha-n)/\beta\delta}f_f^{(n)}(z). \label{genchieq}
\end{equation}
In Ref. \cite{Friman:2011pf}, the scaling behavior of the non-analytic parts of the fourth and sixth order cumulants were found out using $O(4)$ scaling functions. These are shown in Fig.~\ref{o4plots}. Unlike $Z(2)$ universality class, the peak of the fourth order fluctuation is finite at the chiral limit while the sixth order is divergent in both $O(4)$ and $Z(2)$.

From the definition of $t$ in Eq.~\ref{thdef}, it can be easily shown that at $\mu=0$, the following relation holds for some quantity $A$ in the vicinity of the critical point at $T=T_c$,
\begin{equation}
 2\kappa_2^X T\frac{\partial}{\partial T} A = T^2 \frac{\partial^2}{\partial \mu^2}A. \label{scaleexp}
\end{equation}
If the singular part dominates, one expects that a second order fluctuation $\chi_2^X$ would behave like the energy density and similarly, $\chi_4^X$ as the specific heat. This scaling expectation will hold towards the chiral limit as the singular parts dominate.

\begin{wrapfigure}{r}{0.4\textwidth}
  \begin{center}
    \includegraphics[width=.4\textwidth]{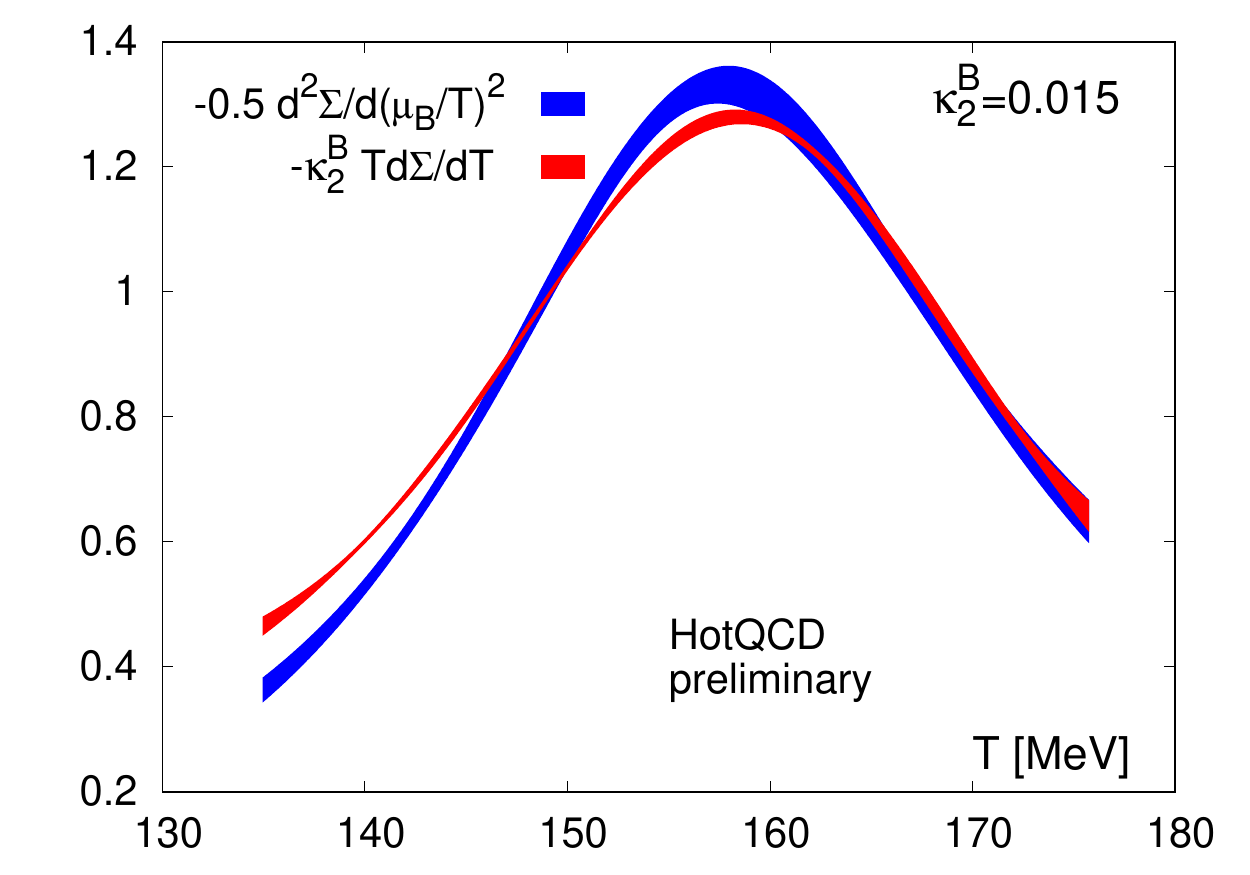}
  \end{center}
  \caption{Dimensionless temperature derivative of the chiral order parameter, $-T\frac{\partial\Sigma}{\partial T}$ compared to the second order Taylor coefficient, $-\frac{1}{2}\frac{\partial^2 \Sigma}{\partial (\mu_B/T)^2} $, in an expansion in $\mu_B$, as a function of temperature. Results have been obtained with lattice temporal extent $N_\tau=12$.} 
\label{pbpder}
\end{wrapfigure}

An example of the scaling expectation can be seen to be valid already at physical quark masses. Let us consider the subtracted chiral condensate\footnote{See Eq.~1 of Ref.~\cite{Bazavov:2018mes} for the definition of $\Sigma$.} $\Sigma$ at $\mu_B=0$  shown in the left most plot of Fig.~1 in Ref.~\cite{Bazavov:2018mes}. From the interpolated curves in the above-mentioned figure, a temperature derivative can be taken, and compared to the second order Taylor coefficient of $\Sigma$ in $\mu_B$, computed in Ref. \cite{Bazavov:2018mes}. In Fig.~\ref{pbpder}, the temperature derivative has been scaled by a factor of $\kappa^B_{2,\mathrm{phys}}=0.015$, the curvature coefficient of the pseudocritical line, determined for physical pion mass in Ref. \cite{Bazavov:2018mes}. 
As seen in the figure, this matches quite well with the second order $\mu_B$ derivative, in terms of the shape of the fluctuations in the whole $T$ range and also, for relative values at the peak positions. This shows that the singular parts are dominant even at physical quark masses for the chiral order parameter. This motivates one to look for observables with dominant criticality which can also be tested experimentally in principle. It also appears that the slope of the transition line does not change too much with quark mass, which has important consequences for the critical point at finite $\mu$ \cite{Karsch:2019mbv}.  

The measurements of the conserved charge fluctuations at smaller-than-physical quark masses have been performed on gauge ensembles generated by the HotQCD collaboration, using $2+1$ flavor highly improved staggered quark (HISQ) and tree-level Symanzik-improved gauge action. These gauge configurations were recently used in the determination of the chiral phase transition temperature $T_c$ in Ref. \cite{Ding:2019prx}. Along with the results at physical mass, which corresponds to $m_l=m_s/27$, we have shown results for $m_l=m_s/40$ and $m_s/80$, all the while keeping the strange quark mass $m_s$ fixed at its physical value. These correspond to pion masses $140$ (physical), $110$ and $80$ MeV, respectively. The current study has been done using the largest available volumes for each mass with temporal extent $N_\tau=8$. We plan to do a comprehensive study in the future with proper continuum and thermodynamic limits.    

\section{Results}
\begin{figure}
\centering
\includegraphics[width=.45\textwidth]{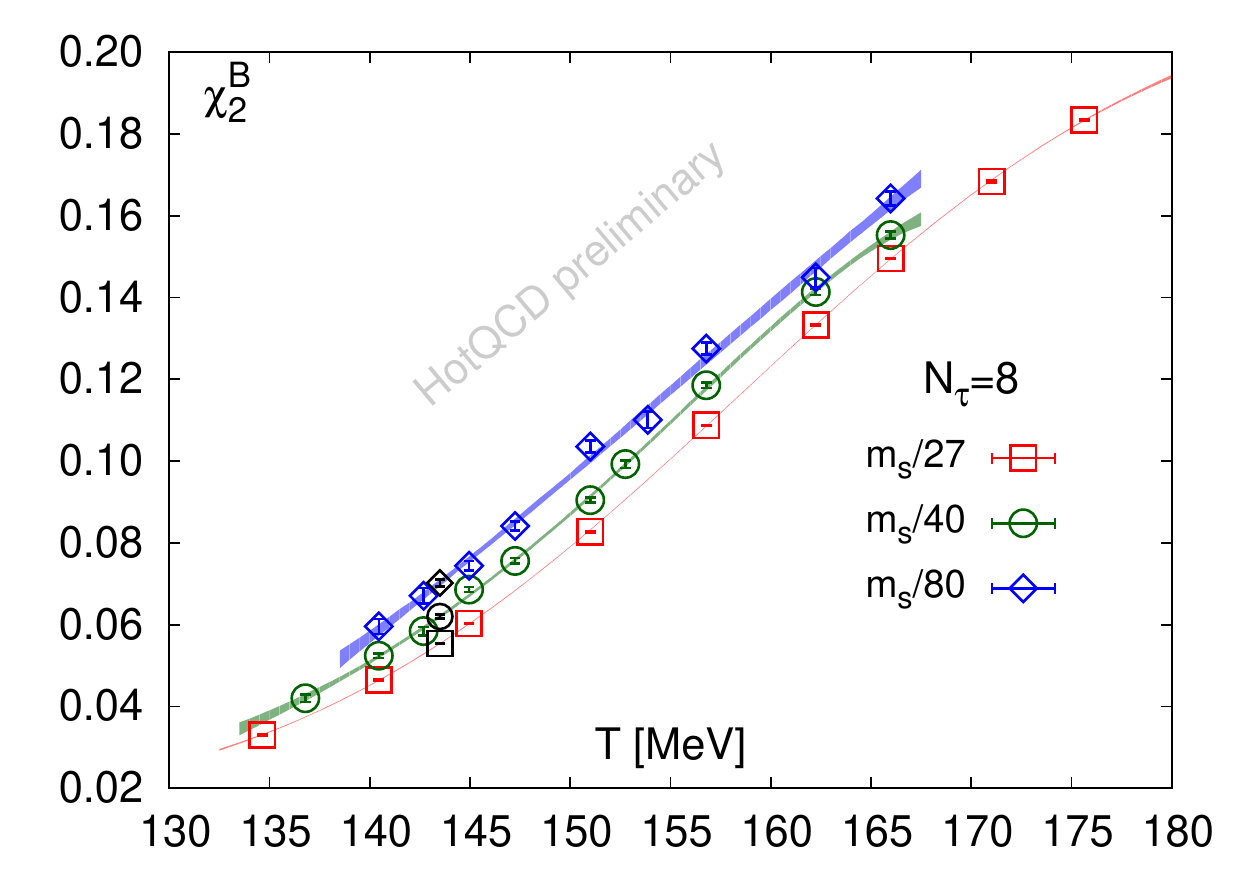}%
\includegraphics[width=.45\textwidth]{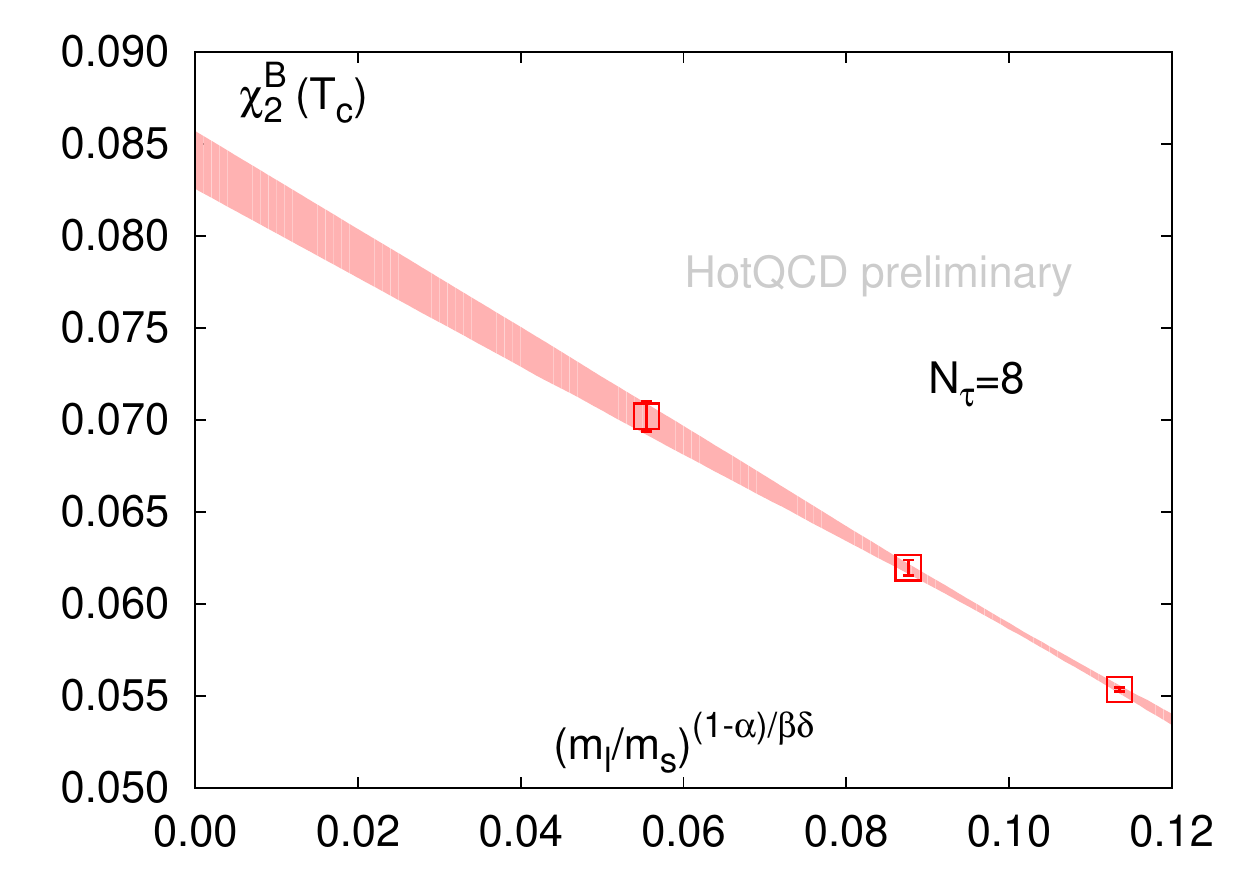}
\caption{Left : The second order baryon number fluctuation as a function of temperature for three different values of $m_l$.
The black symbols indicate the value of $\chi_2^B$ at $T=T_c$. Right : $\chi_2^B(T_c)$ plotted as a function of
$(m_l/m_s)^{(1-\alpha)/\beta\delta}=(m_l/m_s)^{0.662}$, with $O(4)$ critical exponents. The band shows the result of a linear fit.}
\label{chi2b}
\end{figure}

\begin{figure}
\centering
\includegraphics[width=.45\textwidth]{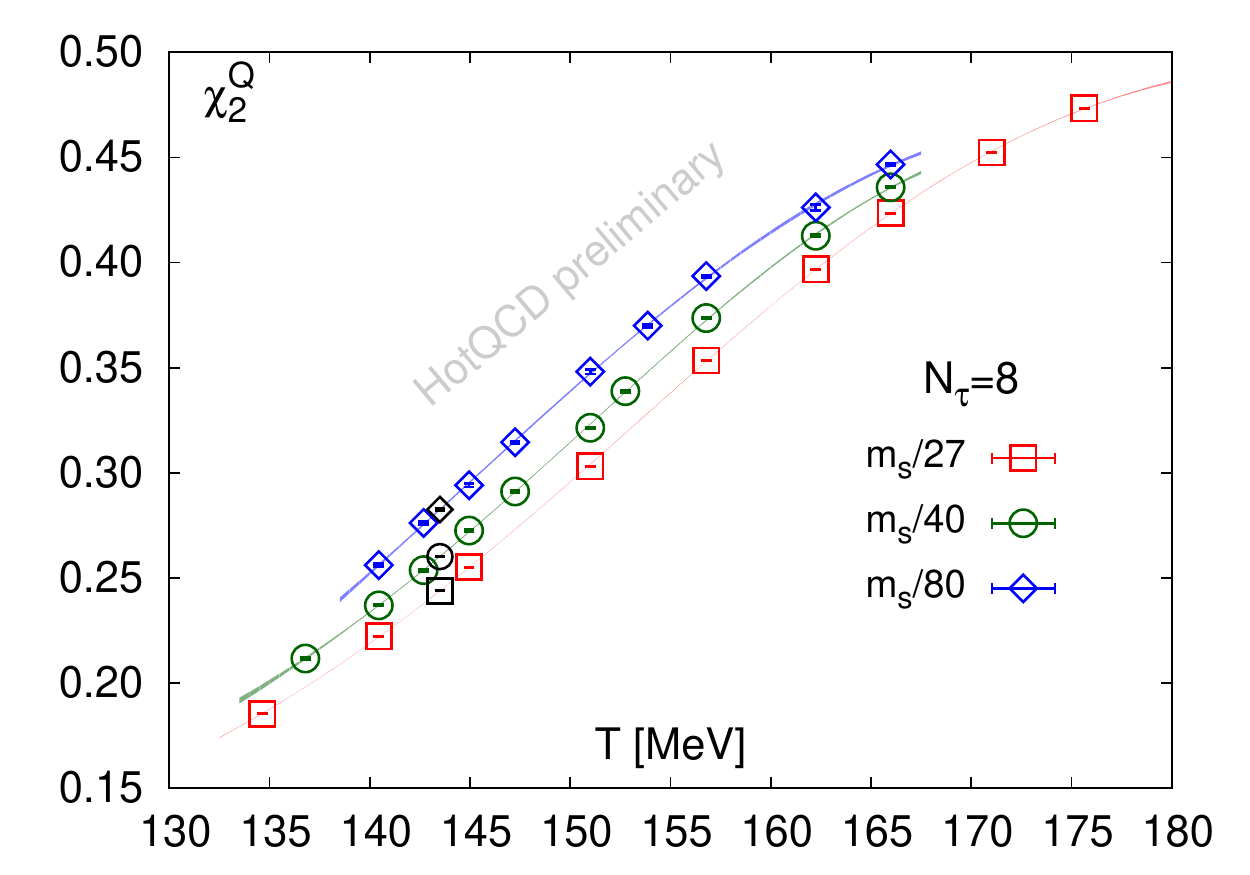}%
\includegraphics[width=.45\textwidth]{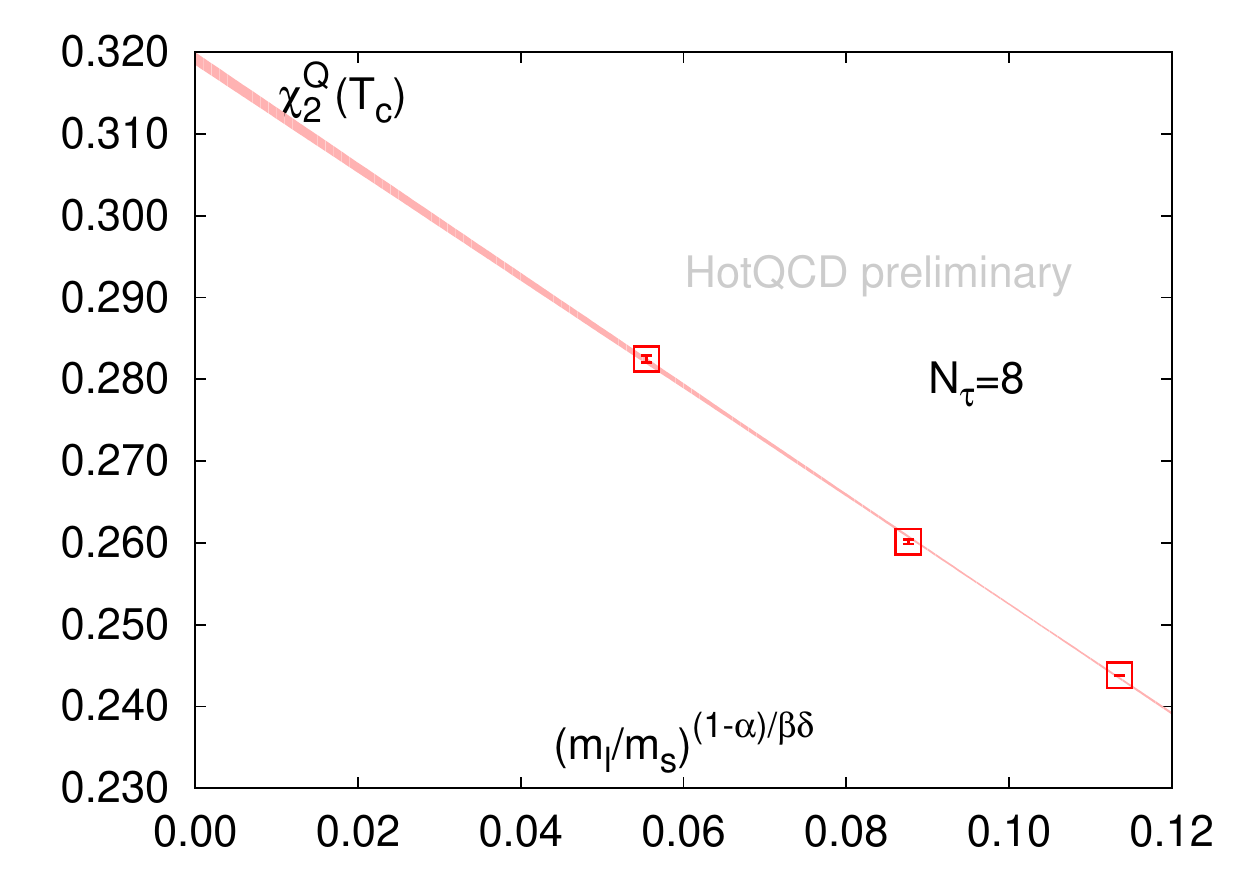}
\includegraphics[width=.45\textwidth]{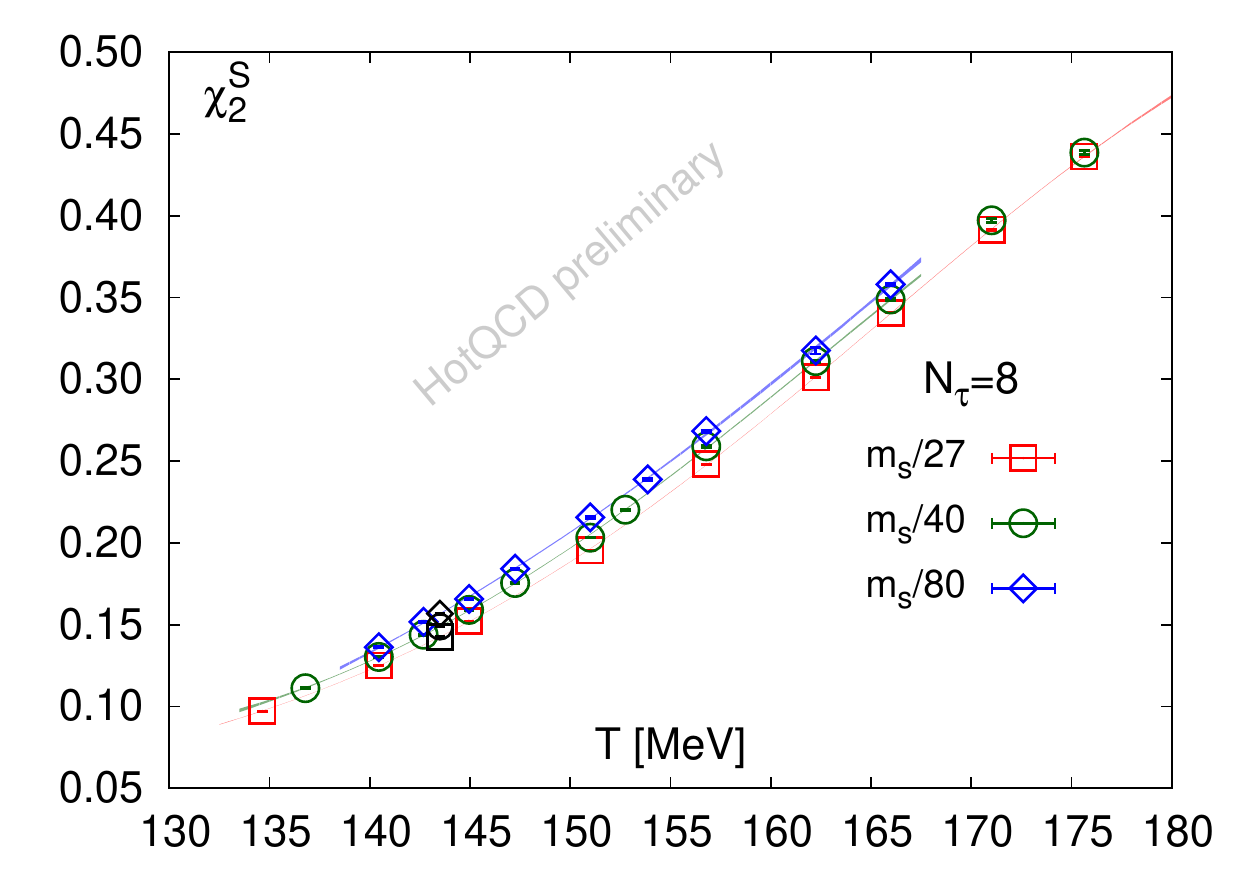}%
\includegraphics[width=.45\textwidth]{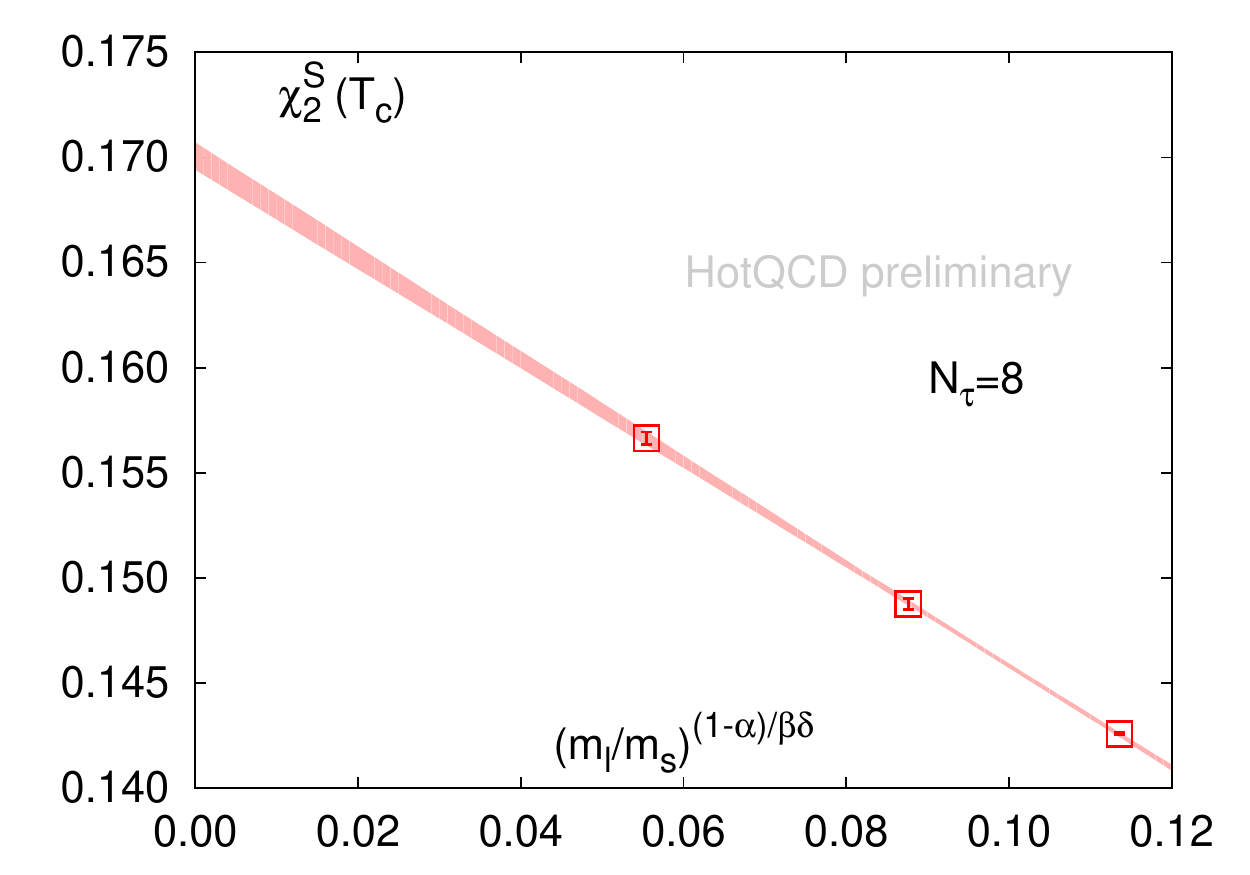}
\caption{Same as Fig.~\ref{chi2b} but for electric charge $Q$ (top) and for strangeness $S$ (bottom).}
\label{chi2q}
\end{figure}

The second order fluctuations, $\chi_2^X$, $X=B,Q,S$ are shown in the left panels of Figs.~\ref{chi2b} and \ref{chi2q}, respectively. They behave qualitatively as an energy density as expected from Eq.~\ref{scaleexp}. However, the regular parts are considerable and therefore, Eq.~\ref{scaleexp} is not exactly satisfied. The singular contribution to second order cumulants of conserved charge fluctuations at physical quark masses can be estimated by following the analysis in Ref.~\cite{Engels:2011km}. Along with the regular terms, the second order fluctuation at $T=T_c$ and $\mu_X=0$ ($t=z=0$) can be written as (cf.~Eq.~\ref{genchieq})
\begin{equation}
 \chi_2^X(T_c,m_l) \sim -2\kappa^X_2 h^{(1-\alpha)/\beta\delta}f_f^{(1)}(0) + C_\mathrm{reg}^X + \mathcal{O}_\mathrm{reg}^X (m_l^2), \label{chi2singular}
\end{equation}
where $C_\mathrm{reg}^X$ is the leading order constant regular term independent of $m_l$.
A linear term in $m_l$ is not allowed from symmetry arguments\footnote{A mass derivative of
the free energy density equivalent to magnetization must flip sign when the mass
(similar to a magnetic field) is reversed.}. 
The values of $\chi_2^B$ extracted at $T=T_c$ for $N_\tau = 8$ \cite{Ding:2019prx}, are shown with black symbols in the left plot in Fig.~\ref{chi2b}.
These values fit quite well with a linear fit in $(m_l/m_s)^{(1-\alpha)/\beta\delta}$ using $O(4)$ critical exponents, shown in the right plot of Fig.~\ref{chi2b}. Strictly speaking, the relevant symmetry group for staggered fermions at finite lattice spacing is $O(2)$, which becomes $O(4)$ only in the continuum limit. However, the exponents of $O(4)$ and $O(2)$ are very close which would result only in a marginal change in the analyses. 

The $y$-intercept gives us the constant regular part since all other terms go to zero.
From the linear fit, one can then estimate the singular part at physical quark masses,
which is just $\chi_2^X (T_c,m_l=0) - \chi_2^X(T_c,m_l=m_s/27)$.
The fact that this fit was done using the critical exponents of $O(4)$ gives
further evidence that the chiral phase transition in $2+1$ flavor QCD is consistent with belonging to the $O(4)$ universality class.

Before moving on to the similar analyses for other conserved charges, it is worth to discuss a subtlety.
Electric charge (chemical potential) breaks the isospin symmetry explicitly, rendering the chiral
symmetry group to be O(2) instead of O(4). Although as mentioned earlier, the effect on the analyses is expected to be marginal. In Fig.~\ref{chi2q}, we show the same analyses for $\chi_2^Q$ (top row) and $\chi_2^S$ (bottom row), with similar findings. 

Using the above analyses, one can also estimate the ratio of the curvature of the critical lines as follows. From Eq.~\ref{chi2singular},
one can see that the ratio of the singular parts of two different second order fluctuations is exactly equal to the ratio of
the corresponding curvature coefficients. 
For example after estimating the singular parts from the fits in Fig.~\ref{chi2b} and \ref{chi2q}, we find 
\begin{equation}
 \frac{\textrm{Singular part of }\chi_2^S}{\textrm{Singular part of }\chi_2^B} = \frac{\kappa_2^S}{\kappa_2^B} \sim 1,
\end{equation}
which agrees with the conservative estimate, $1.1 \pm 0.5$, computed from the coefficients obtained at physical pion mass \cite{Bazavov:2018mes}.
Also, the ratio of $\kappa_2^Q$ and $\kappa_2^B$ is roughly consistent with previous results \cite{Bazavov:2018mes}.
This indirectly shows that the slope of the transition line does not change much with mass (cf. Fig.~\ref{3dphasediag}).
It is worth pointing out here that the baryon number fluctuations require more statistics compared to electric charge
and strangeness fluctuations as is evident from the Figs.~\ref{chi2b} and \ref{chi2q}.

\begin{figure}
\centering
\includegraphics[width=.45\textwidth]{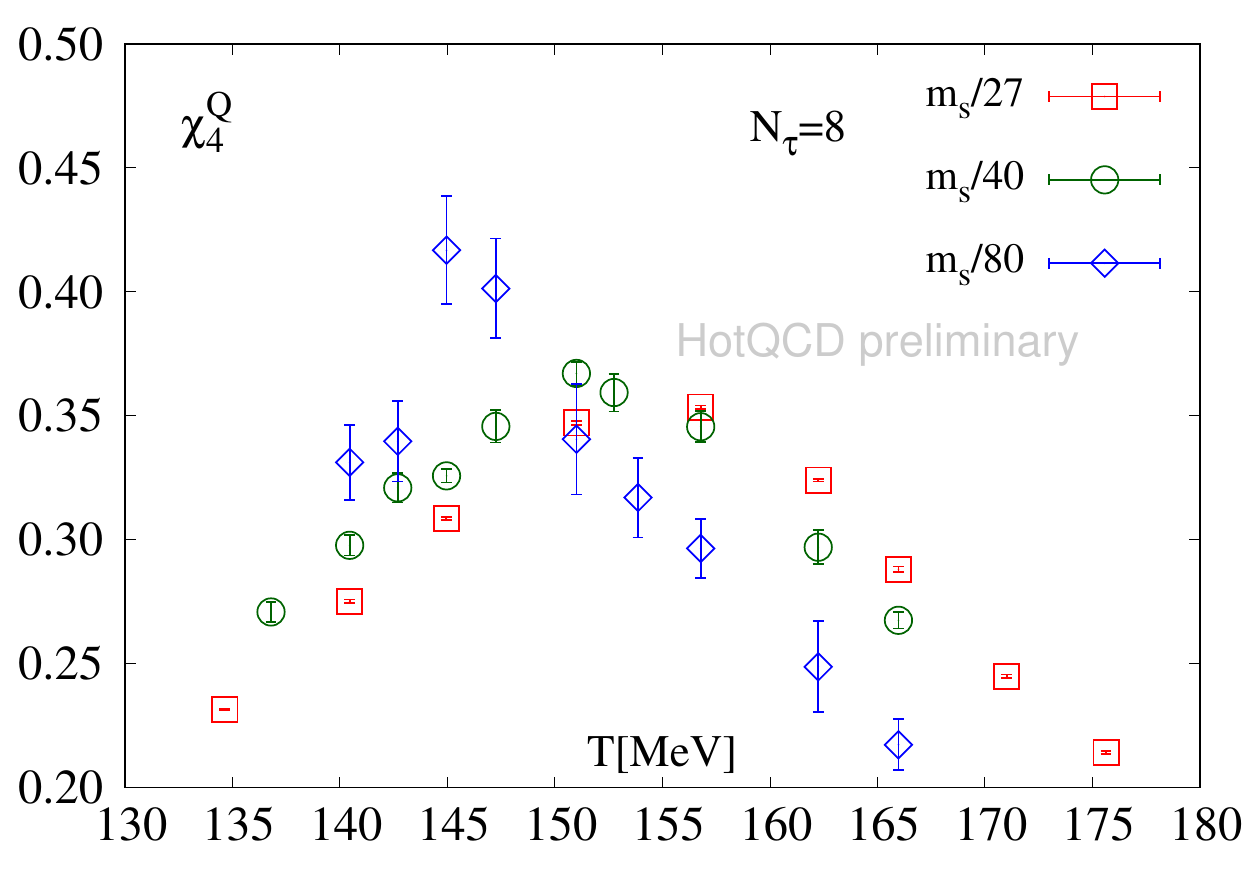}~~~
\includegraphics[width=.45\textwidth]{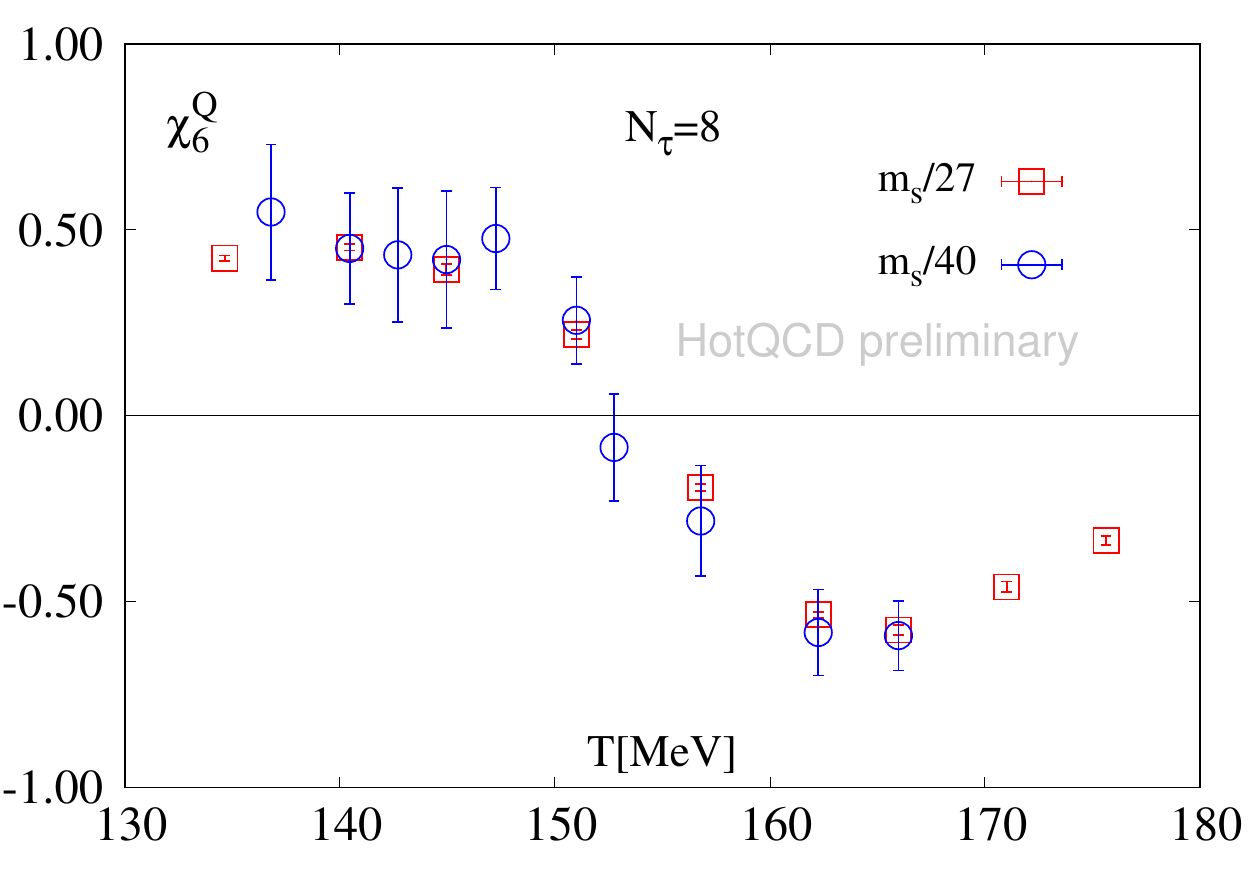}
\caption{Left : The fourth order fluctuation $\chi_4^Q$ as a function of temperature for different masses. Right : The sixth order fluctuation $\chi_6^Q$ as a function of temperature for different masses.}
\label{chi4chi6plot}
\end{figure}

The critical behavior of fourth order fluctuations of the free energy density w.r.t.\ chemical potential is expected to be like the specific heat. It was shown in Ref.~\cite{Friman:2011pf} that the universal part of the specific heat goes to zero as $h^{0.116}f_f^{(2)}$ with $h\to 0$, according to 3d $O(4)$ universality class (see left plot in Fig.~\ref{o4plots}). If the peak of the fourth order fluctuation saturates at the chiral limit following the same mass scaling, one can confirm the universality class of the chiral phase transition. Keeping in mind the regular contribution, the fourth order fluctuation $\chi_4^Q$, shown in Fig.~\ref{chi4chi6plot} (left), is consistent with the $O(4)$ universal function shown in the left plot in Fig.~\ref{o4plots}. The behavior is different from $Z(2)$ scaling expectation where $\chi_4$ is expected to diverge in the chiral limit.    

The sixth order fluctuations is expected to diverge at the chiral limit according to both $O(4)$ and $Z(2)$ universality classes.
The mass scaling of the peaks will determine the universality class. For the sixth order fluctuations,
below the crossover temperature $T_{pc}$ at any quark mass, there will be considerable regular contribution
as expected from HRG calculations. So, the interesting region to look at will be the region above $T_{pc}$ which is
expected to be dominated mostly by the singular part. As seen in the right plot in Fig.~\ref{o4plots}, the minimum of the
universal scaling function decreases as $h^{-0.429}$ \cite{Friman:2011pf}. As a result, the ratio of the peak heights
between $m_l=m_s/40$ and $m_l=m_s/27$ expected from $O(4)$ scaling is $\sim 1.18$.
In the right plot of Fig.~\ref{chi4chi6plot}, we have plotted preliminary results for the sixth order electric charge fluctuation
for the above-mentioned mass ratios. Although one has to keep in mind that electric charge fluctuations contain the isospin
fluctuations as a regular background which could mask the singular contribution stemming from baryon number fluctuations.
The results for baryon number fluctuations (except second order) are still quite noisy and require more statistics.

\section{Conclusions}
To summarise, our preliminary results of the conserved charge fluctuations indicate the existence of a second order chiral phase transition with 3d $O(4)$ universality class in the chiral limit. We demonstrate a way to estimate the singular and regular contributions of the second order fluctuations. Subsequently, the ratios of the curvature coefficients at chiral limit obtained from the ratios of the singular parts, show them to be consistent with similar ratios at physical pion mass. We are currently working to make these results concrete with more statistics and proper continuum limit.

\end{document}